\documentclass[letterpaper,10pt]{article} 

\usepackage{opticameet3} 

\newcommand\authormark[1]{\textsuperscript{#1}}

\usepackage{amsmath,amssymb}
\usepackage{algorithm}
\usepackage{algpseudocode}
\usepackage{booktabs}
\usepackage{url}
\usepackage{graphicx}
\usepackage[colorlinks=true,bookmarks=false,citecolor=blue,urlcolor=blue]{hyperref} 

\newtheorem{definition}{Definition}

\begin{document}






\title{An information-geometric approach for network decomposition using the q-state Potts model}

\author{Alexandre L. M. Levada\authormark{1}}

\address{Computing Department, Federal University of S\~ao Carlos, S\~ao Carlos, SP, Brazil}


\email{alexandre.levada@ufscar.br} 

\begin{abstract}
Complex networks are critical in many scientific, technological, and societal contexts due to their ability to represent and analyze intricate systems with interdependent components. Often, after labeling the nodes of a network with a community detection algorithm, its modular organization emerges, allowing a better understanding of the underlying structure by uncovering hidden relationships. In this paper, we introduce a novel information-geometric framework for the filtering and decomposition of networks whose nodes have been labeled. Our approach considers the labeled network as the outcome of a Markov random field modeled by a q-state Potts model. According to information geometry, the first and second order Fisher information matrices are related to the metric and curvature tensor of the parametric space of a statistical model. By computing an approximation to the local shape operator, the proposed methodology is able to identify low and high information nodes, allowing the decomposition of the labeled network in two complementary subgraphs. Hence, we call this method as the LO-HI decomposition. Experimental results with several kinds of networks show that the high information subgraph is often related to edges and boundaries, while the low information subgraph is a smoother version of the network, in the sense that the modular structure is improved.
\end{abstract}

\section{Introduction}

The motivation for studying complex networks in modern science is driven by the need to understand and optimize systems characterized by intricate interconnections and interactions, which are prevalent in both natural and engineered environments \cite{CN1}. Complex networks provide a powerful framework for modeling a wide array of phenomena, from biological processes and social interactions to economic dependencies and climate change \cite{CN2}. This field offers critical insights into the structural properties, dynamics, and robustness of systems, enabling the prediction and control of behaviors such as information dissemination, disease spread, and network failures \cite{CN3}. Additionally, network analysis supports the optimization of resources and efficiency in applications like transportation, communication, and supply chains. With the advent of big data, network theory facilitates the extraction of meaningful patterns from vast datasets, enhancing the capabilities of machine learning and AI \cite{CN4}. Furthermore, it plays a crucial role in addressing global challenges such as cybersecurity by informing targeted interventions, policy decisions and public health \cite{CN5}. Thus, the study of complex networks is essential for advancing our understanding of interconnected systems and developing innovative solutions to contemporary scientific and societal issues.

Markov Random Fields (MRFs) are crucial in the study of complex networks due to their ability to model the dependencies between interconnected variables in a probabilistic framework \cite{MRF1}. MRFs provide a robust method for representing the local interactions within a network, capturing the conditional dependencies that define the overall system behavior \cite{MRF2}. This capability is particularly important in domains such as image processing, where MRFs are used to model pixel relationships for tasks like image segmentation and denoising \cite{MRF3}, and in social network analysis, where they help understand the influence dynamics and community structures \cite{MRF4}. By leveraging the local independence properties of MRFs, researchers can efficiently infer global properties of the network, perform tasks such as prediction \cite{MRF5} and anomaly detection \cite{MRF6}, and optimize the network configuration.

The Potts model, a generalization of the Ising model, is a MRF widely used in the study of complex networks, particularly in the context of statistical physics, image processing, and spatial data analysis \cite{Potts1}. The Potts MRF model is characterized by a set of nodes, each assigned one of a finite number of discrete states or labels, with edges representing interactions between neighboring nodes \cite{Potts2}. The model assigns a higher probability to configurations where neighboring nodes have the same state, thus promoting local uniformity and capturing the tendency of similar nodes to cluster together at low temperatures \cite{Potts3}. This property makes the Potts model especially useful for tasks such as image segmentation, where it helps in identifying homogeneous regions, and in community detection in networks, where it identifies groups of densely connected nodes. The energy function of the Potts model includes terms for both node-specific potentials and pairwise interactions, allowing it to flexibly model various types of dependencies and interactions within the network \cite{Potts4}.

One question that often arises among researchers and practitioners in network science is: what kind of analysis can we perform after community detection in complex networks? In this paper, we show a possible direction to this open question by proposing an information-geometric approach for network filtering and decomposition based in the q-state Potts model to detect the most informative nodes in labeled graphs. The main idea here is to consider networks with their detected communities as outcomes of pairwise isotropic Potts Markov random field models and use information geometry to extract high information patterns from these networks.

Recently, information geometry results have shown that the Fisher information matrix of statistical model are directly related to the metric and curvature tensors of its parametric space \cite{Amari2021,Dodson,Nielsen1,Nielsen2}. In this paper, by using both first and second order Fisher information expressions, we propose an approximation to the local shape operator as a way to measure the amount of information of each configuration patch of the network (i.e., the set defined by a node and its neighbors). Using this definition, we can filter and decompose a given network into two complementary disjoint components: a low information subgraph and a high information subgraph. Our results indicate that there is a close relation between the high information subgraph, edges, boundaries and long-range interactions in labeled networks. Moreover, the high information subgraph of a network is mainly composed by local configuration patterns that does not fit the expected global behavior, encoded by the inverse temperature parameter (smaller values means noisy behavior and larger values means smooth behavior).

In summary, the main contributions of this paper are twofold: 1) we derive mathematical expressions for the computation of the first and second order Fisher information in the pairwise isotropic q-state Potts model and, 2) we propose a network decomposition strategy based in low and high information subgraphs, which is useful for filtering networks based on the labels of their nodes. Examples with several network topologies show that, in smooth labelings, it is possible to increase modularity of communities in the low information subgraph, suggesting that this representation resembles a smoother version of the original network. Moreover, in smooth labelings, the analysis of high information subgraphs indicate that this representation encodes edges and boundaries, playing an important role in the spreading information along communities due to long range interactions.  

The remaining of the paper is organized as follows: Section 2 briefly describes the q-state Potts MRF model. Section 3 presents the maximum pseudo-likelihood technique for the estimation of the inverse temperature parameter. Section 4 presents Fisher information and Section 5 describes the mathematical derivations of the first and second order expressions for the q-state Potts modeland the proposed information-geometric network decomposition method based in the q-state Potts model. Section 6 shows our computational experiments and results. Finally, Section 7 presents our conclusions and final remarks.

\section{The isotropic q-State Potts model}

The q-state Potts model is a discrete Markov model, in the sense that the variables belonging to the random field can only assume discrete values from a finite set $C = \{ 1,2, ..., q \}$. It is a model that tries to imitate the way in which individual elements of the most diverse types (i.e., atoms, animals, biological structures, individuals of a society, etc.) modify their behavior according to the behavior of other elements belonging to the context in which they are found (neighborhood). In other words, it is a model widely used to study collective effects as a consequence of local interactions. It is one of the most widely known Markovian models to date. References in literature indicate that the Potts model plays a fundamental role in research of the most varied areas, such as mathematics \cite{PottsMath1,PottsMath2,PottsMath3,PottsMath4}, physics \cite{PottsPhys1,PottsPhys2,PottsPhys3}, biology \cite{PottsBiol1,PottsBiol2,PottsBiol3}, computing \cite{PottsCS1,PottsCs2} and even sociology \cite{PottsSoc}.

According to the Hammersley-Clifford Theorem \cite{Hammersley}, the q-state Potts model can be equivalently defined in two different ways: by a joint distribution (Gibbs random field) or by a set of local conditional density functions (Markov random field). For the purposes of this work, representation by local models is more appropriate as it allows the direct calculation of probabilities for node of the field individually and not for an entire outcome of the random field as would be the case from the Gibbs distribution. Considering a general neighborhood system $\eta_{i}$, the local conditional density function of a pairwise isotropic q-state Potts model is \cite{LCDF}:

\begin{equation}
	p\left( x_{i} = m | \eta_{i}, \beta \right) = \frac{exp\{ \beta U_{i}(m) \}}{\displaystyle \sum_{\ell=1}^{q}exp\{ \beta U_{i}(\ell) \}}
\end{equation} where $U_{i}$ denotes the number of neighboring nodes having the same label as the central node $x_{i}$, $q$ is the number of states (labels) and $\beta$ is the inverse temperature parameter, which controls the global dependency structure. It is worth mentioning that the higher the value of $\beta$, the smoother the random field outcomes, as the most likely configurations have a great degree of agreement between neighboring nodes. Note however that, when $\beta = 0$, that is, the nodes are statistically independent, the model degenerates to the probability mass function of a discrete random variable with uniform distribution (all states are equiprobable) and the contextual information is completely ignored. A particular case of the Potts model, when $q = 2$, is the classical Ising model \cite{Ising,Ising2}, used in statistical mechanics to study the behavior of physical systems of many particles. Basically, in this model, the system can present two well-defined types of behavior: a ferromagnetic, a situation in which spin alignment prevails (low temperature favors low energy configurations), or anti-ferromagnetic, when spin misalignment prevails (high temperatures).

\section{Maximum pseudo-likelihood estimation}

One of the greatest difficulties in Markov random field models is the estimation of the coupling parameters. Basically, this is due to the fact that the most widely known technique, maximum
likelihood (ML), cannot be applied directly because the presence of the partition function in the joint Gibbs distribution, which is computationally intractable \cite{BesagMPL}. An alternative solution, consists of using the local conditional density functions in maximum pseudolikelihood (MPL) estimation \cite{BesagMPL}. Basically, the main motivations for using this approach are:

\begin{itemize}
	\item It is a computationally viable method.
	\item From a statistical point of view, MPL estimators have a series of desirable properties, such as consistency and asymptotic normality \cite{MPL}. Thus, its behavior can be completely characterized the limiting case.
\end{itemize}

The pseudo-likelihood function for the q-state Potts model is defined as:

\begin{equation}
	PL(\beta) = \prod_{i=1}^{n} p\left( x_{i} | \eta_{i}, \beta \right) = \prod_{i=1}^{n} \left[ \frac{exp\{ \beta U_{i}(x_i) \}}{\displaystyle \sum_{\ell=1}^{q}exp\{ \beta U_{i}(\ell) \}} \right]
\end{equation} where $n$ denotes the number of nodes in the network. Due to tractability issues, it is often recommended to maximize the log-likelihood, which is given by:

\begin{align}
	log~PL(\beta) = \sum_{i=1}^{n} \left[ \beta U_{i}(x_i) - \sum_{\ell=1}^{q}exp\{ \beta U_{i}(\ell) \}  \right]
\end{align}

By differentiating the log-pseudo-likelihood equation w.r.t. the inverse temperature parameter and setting the result to zero, we have:

\begin{align}
	\frac{\partial}{\partial\beta}log~PL(\beta) = \sum_{i=1}^{n} U_i(x_i) - \sum_{i=1}^{n}\left[ \frac{\displaystyle\sum_{\ell=1}^{q}U_i(\ell)exp\{\beta U_i(\ell)\}}{\displaystyle\sum_{\ell=1}^{q}exp\{\beta U_i(\ell)\}} \right] = 0
\end{align} where the first term is the free energy and the second term is the expected energy. In other words, the MPL estimative is the value of inverse temperature for which the free energy of the system is equal to the expected energy. The previous equation can be solved by a numerical root finding algorithm. In this paper, we employ the secant method, since it does not require the computation of the derivative of the original function.

\section{Fisher information}

Information geometry is a field that merges differential geometry with information theory, focusing on the geometric structures and properties of probability distributions. It provides a powerful framework for understanding and analyzing statistical models by treating them as geometric objects in a high-dimensional space \cite{Amari2021,Dodson,Nielsen1,Nielsen2}. Within this context, the Fisher information metric plays a pivotal role as it defines the Riemannian metric on the space of probability distributions. Fisher information measures the amount of information that an observable random variable carries about an unknown parameter upon which the probability depends. It provides a way to quantify the sensitivity of a probability distribution to changes in its parameters, essentially capturing the "curvature" of the statistical manifold. This curvature reflects how easily distinguishable nearby probability distributions are, thereby influencing the efficiency of estimators and the behavior of learning algorithms. In information geometry, the Fisher information matrix serves as the metric tensor, allowing for the application of geometric tools and concepts, such as geodesics and curvature, to the study of statistical inference and machine learning. This geometric perspective aids in developing more robust statistical methods, understanding the limitations of estimators, and optimizing learning algorithms in complex, high-dimensional spaces. In summary, given a random variable with probability density function $p(x; \theta)$, the first-order Fisher information matrix is the metric tensor of the parametric space, while the second-order Fisher information matrix encodes information about the curvature of the parametric space. Briefly speaking, information geometry is a framework enables the macroscopic description of random systems to be done using the same mathematical instruments that are used to analyze deterministic physical phenomena, such general relativity \cite{Quantum}.

\begin{definition}[First-order Fisher information]
	Let $p(X;\theta)$ be a probability density function where $\theta \in \Theta$ is the parameter of the model. The first-order Fisher information is defined as \cite{Fisher}:
	\begin{equation}
		\mathbb{I}(\theta) = \mathbb{E}\left[ \left(\frac{\partial}{\partial\theta} log~p(X; {\theta}) \right) \right]^2
	\end{equation}
\end{definition}

\begin{definition}[Second-order Fisher information]
	Let $p(X;\theta)$ be a probability density function where $\theta \in \Theta$ is the parameter of the model. The second-order Fisher information is defined as \cite{Fisher}:
	\begin{equation}
		\mathbb{II}(\theta) = -\mathbb{E} \left[\frac{\partial^2}{\partial\theta^2} log~p(X; {\theta}) \right] 
	\end{equation}
\end{definition}

Basically, Fisher information is a widely known measure in statistical inference of models belonging to the exponential family under the assumption of independence. In summary, it provides the amount of information that a i.i.d sample carries in relation to an unknown parameter. Under certain regularity conditions, it has been shown that the information equality holds, which
means that the first and second-order Fisher information are identical. However, in the q-state Potts model, this is not the case.

\section{The proposed LO-HI decomposition method}

In this section, we derive the mathematical expressions for the Fisher information in the q-state Potts model and propose the information-theoretic network decomposition strategy.

\subsection{Fisher information in the q-state Potts model}
 
The observed Fisher information allows the approximation of the expectations in both first and second-order Fisher information, by the law of the large numbers \cite{ObsFisher}. In summary, we can approximate the first-order Fisher information $\mathbb{I}(\beta)$ as an average:

\begin{align}
	\mathbb{I}(\beta) \approx \Phi({\beta}) = \frac{1}{n}\sum_{i=1}^{n}\left[ \frac{\partial}{\partial\beta} \log p(x_i|\eta_i, \hat{\beta}) \right]^2
\end{align} where $\hat{\beta}$ denotes an estimative for the inverse temperature parameter. In a similar way, it is possible to approximate the second-order Fisher information $\mathbb{II}(\beta)$ as:

\begin{align}
	\mathbb{II}(\beta) \approx \Psi({\beta}) = -\frac{1}{n}\sum_{i=1}^{n} \frac{\partial^2}{\partial\beta^2} \log p(x_i|\eta_i, \hat{\beta})
\end{align}

Note that these approximations are computed as the sample averages of the local Fisher information for each node of the network, denoted by $\Phi_i({\beta})$ and $\Psi_i({\beta})$:

\begin{align}
	\Phi_i({\beta}) & = \left[ \frac{\partial}{\partial\beta} \log p(x_i|\eta_i, \hat{\beta}) \right]^2 \\ 
	\Psi_i({\beta}) & = - \frac{\partial^2}{\partial\beta^2} \log p(x_i|\eta_i, \hat{\beta})
\end{align}

We begin with the derivation of $\Phi_i({\beta})$, by plugging equation (1) into equation (9):

\begin{align}
	\Phi_i({\beta}) & = \left[ U_i(x_i) - \left( \frac{\displaystyle\sum_{\ell=1}^{q}U_i(\ell)exp\{\beta U_i(\ell)\}}{\displaystyle\sum_{\ell=1}^{q}exp\{\beta U_i(\ell)\}} \right) \right]^2 
\end{align}

By grouping the first and second terms, the equation becomes:

\begin{align}
	\Phi_i({\beta}) & = \left[ \frac{\displaystyle \sum_{\ell=1}^{q}U_i(x_i)exp\{\beta U_i(\ell)\} - \sum_{\ell=1}^{q}U_i(\ell)exp\{\beta U_i(\ell)\}}{\displaystyle\sum_{\ell=1}^{q}exp\{\beta U_i(\ell)\}} \right]^2 = \left[ \frac{\displaystyle \sum_{\ell=1}^{q}(U_i(x_i) - U_i(\ell))exp\{\beta U_i(\ell)\}}{\displaystyle\sum_{\ell=1}^{q}exp\{\beta U_i(\ell)\}} \right]^2
\end{align}

Finally, by expanding the square, we have:

\begin{align}
	\Phi_i({\beta}) = \frac{\displaystyle \sum_{\ell=1}^{q}\sum_{k=1}^{q} (U_i(x_i) - U_i(\ell))(U_i(x_i) - U_i(k))exp\{\beta (U_i(\ell)+U_i(k))\}}{\displaystyle\sum_{\ell=1}^{q}\sum_{k=1}^{q}exp\{\beta (U_i(\ell) + U_i(k))\}} 
\end{align}

To compute $\Psi_i({\beta})$, first note that:

\begin{align}
	\Psi_i({\beta}) = \frac{\displaystyle\left[ \sum_{\ell=1}^{q}U_i(\ell)^2 exp\{\beta U_i(\ell)\} \right] \left[ \sum_{\ell=1}^{q}exp\{\beta U_i(\ell)\}\right] - \left[ \sum_{\ell=1}^{q}U_i(\ell)exp\{\beta U_i(\ell)\}\right]}{\displaystyle\left[ \sum_{\ell=1}^{q}exp\{\beta U_i(\ell)\}\right]^2}
\end{align} which, after some algebra, can be expressed as:

\begin{align}
	\Psi_i({\beta}) = \frac{\displaystyle\left[ \sum_{\ell=1}^{q}\sum_{k=1}^{q}U_i(\ell)^2 exp\{\beta (U_i(\ell)+U_i(k))\} - \sum_{\ell=1}^{q}\sum_{k=1}^{q}U_i(\ell)U_i(k) exp\{\beta (U_i(\ell)+U_i(k))\} \right]}{\displaystyle\left[ \sum_{\ell=1}^{q}exp\{\beta U_i(\ell)\}\right]^2}
\end{align} and finally leads to:

\begin{align}
	\Psi_i({\beta}) = \frac{\displaystyle \sum_{\ell=1}^{q}\sum_{k=1}^{q}(U_i(\ell)^2 - U_i(\ell)U_i(k)) exp\{\beta (U_i(\ell)+U_i(k))\} }{\displaystyle \sum_{\ell=1}^{q}\sum_{k=1}^{q}exp\{\beta (U_i(\ell) + U_i(k))\}}
\end{align}

It is possible to observe that a condition to the information equality, which means, $\Phi_i({\beta}) = \Psi_i({\beta})$, is:

\begin{equation}
	(U_i(x_i) - U_i(\ell))(U_i(x_i) - U_i(k)) = U_i(\ell)^2 - U_i(\ell)U_i(k)
\end{equation}

By direct computation of the product in the left side of the equality, we have:

\begin{equation}
	U_i(x_i)^2 - U_i(x_i)U_i(k) - U_i(x_i)U_i(\ell) = U_i(\ell)^2 - 2 U_i(\ell)U_i(k)
\end{equation}

Adding $U_i(k)^2$ to both sides leads to:

\begin{equation}
	U_i(x_i)^2 + U_i(k)^2 - U_i(x_i)U_i(k) - U_i(x_i)U_i(\ell) = (U_i(\ell) - U_i(k))^2
\end{equation}

Adding and subtracting $2U_i(x_i)U_i(k)$ from the left side of the equality finally leads to:

\begin{equation}
	(U_i(x_i) - U_i(k))^2 + U_i(x_i)U_i(k) - U_i(x_i)U_i(\ell) = (U_i(\ell) - U_i(k))^2 + U_i(x_i)U_i(\ell)
\end{equation}

As all terms of the equality are non-negative, the only way to satisfy the equality is having: 

\begin{itemize}
	\item $U_i(x_i) = U_i(\ell)$ to make the quadratic terms identical;
	\item $U_i(\ell) = U_i(k)$ to make the products identical.
\end{itemize}

Therefore, by combining the two conditions, the condition for information equality is:

\begin{equation}
	U_i(x_i) = U_i(\ell) = U_i(k)
\end{equation} which means that the frequency of the labels in the neighborhood of each node must be the same (the local distribution of the labels is always uniform). However, if the information equilibrium holds in the q-state Potts model, then the inverse temperature parameter is zero, which means that the random variables in the field are independent.

\subsection{Fisher information in tensorial notation}

In order to speed up the computation of the first and second-order Fisher information in the q-state Potts model, we propose to express the quantities $\Phi_i(\beta)$ and $\Psi_i(\beta)$ in terms of tensorial products. Let the vectors $\vec{v}_i$ and $\vec{w}_i$ be defined as follows:

\begin{align}
	\vec{v}_i = 
	\begin{bmatrix}
		U_i(x_i) - U_i(1)  \\
		U_i(x_i) - U_i(2)  \\
		\vdots					\\
		U_i(x_i) - U_i(q) 
	\end{bmatrix}
	\\ \nonumber \\
	\vec{w}_i = 
	\begin{bmatrix}
		exp\{\beta U_i(1)\}  \\
		exp\{\beta U_i(2)\} \\
		\vdots					\\
		exp\{\beta U_i(q)\} 
	\end{bmatrix}
\end{align} and the $q \times q$ matrices $A_i$ and $B_i$ as:

\begin{align}
	A_i & = 
	\begin{bmatrix}
		U_i(1) & U_i(1) & U_i(1) & ... & U_i(1) \\
		U_i(2) & U_i(2) & U_i(2) & ... & U_i(2) \\
		\vdots & \vdots & \vdots & \ddots & \vdots \\
		U_i(q) & U_i(q) & U_i(q) & ... & U_i(q) 
	\end{bmatrix}
	\\ \nonumber \\
	B_i & = 
	\begin{bmatrix}
		0 & U_i(1) - U_i(2) & U_i(1) - U_i(3) & ... & U_i(1) - U_i(q) \\
		U_i(2) - U_i(1) & 0 & U_i(2) - U_i(3) & ... & U_i(2) - U_i(q) \\
		\vdots & \vdots & \vdots & \ddots & \vdots \\
		U_i(q) - U_i(1) & U_i(q) - U_i(2) & U_i(q) - U_i(3) & ... & 0 
	\end{bmatrix}
\end{align} with $\Lambda_i = A_i \odot B_i$, where $\odot$ denotes the Haddamard (pointwise) product and $\otimes$ denotes the Kronecker product, the expressions for $\Phi_i(\beta)$ and $\Psi_i(\beta)$ can be expressed as:

\begin{align}
	\Phi_i(\beta) & = \frac{\left\lVert \left( \vec{v}_i \odot \vec{w}_i \right) \otimes \left( \vec{v}_i \odot \vec{w}_i \right)^T \right\rVert_{+}}{\left\lVert \vec{w}_i \otimes \vec{w}_i \right\rVert_{+}} \\
	\nonumber \\
	\Psi_i(\beta) & = \frac{\left\lVert \Lambda_i \odot \left( \vec{w}_i \otimes \vec{w}_i \right) \right\rVert_{+}}{\left\lVert \vec{w}_i \otimes \vec{w}_i \right\rVert_{+}}
\end{align} where $\lVert A \rVert_{+}$ denotes the summation of the elements of the matrix/vector $A$. In order to compute the local curvatures at each node, in computational terms, we have to define the local shape operator. The shape operator encodes relevant information about the curvature of surfaces, being a powerful mathematical tool for geometric analysis. From the shape operator, we can obtain the Gaussian, mean and principal curvatures as its determinant, trace and eigenvalues \cite{Manfredo}.

\begin{definition}[Shape operator]
	Let $M$ be a surface with first fundamental form $\mathbb{I}$ and second fundamental form $\mathbb{II}$. Then, the shape operator $P$ can be computed by \cite{Manfredo}:
	\begin{equation}
		P = -\mathbb{II}(\theta)\mathbb{I}(\theta)^{-1} 
	\end{equation}	
\end{definition}

In the q-state Potts model, the parametric space is one-dimensional and the shape operator is a scalar value that measures the curvature of the parametric space at the point $\beta$. Therefore, of particular interest for the proposed network filtering strategy is:

\begin{align}
	S_i(\beta) = -\frac{\Psi_i(\beta)}{\Phi_i(\beta)} = -\frac{\left\lVert \Lambda_i \odot \left( \vec{w}_i \otimes \vec{w}_i \right) \right\rVert_{+}}{\left\lVert \left( \vec{v}_i \odot \vec{w}_i \right) \otimes \left( \vec{v}_i \odot \vec{w}_i \right)^T \right\rVert_{+} + \lambda}
\end{align} where $\lambda = 0.001$ is a small regularization parameter to avoid division by zero.

\subsection{LO-HI decomposition: an information-theoretic network partition}

Graph decomposition is crucial in the analysis of complex networks because it facilitates the understanding of the intricate structural properties and functionalities of these networks \cite{NetDecomp1}. By breaking down a large and often convoluted graph into smaller, more manageable subgraphs, researchers can analyze these components individually and gain insights that are otherwise obscured by the network's complexity \cite{NetDecomp2}. This process helps in identifying key structural features such as communities, clusters, and hierarchical organization within the network \cite{NetDecomp3}. Additionally, graph decomposition aids in optimizing computational tasks by reducing the problem size, enabling more efficient algorithmic processing and scalability. It also plays a significant role in uncovering hidden patterns and relationships, which are essential for applications such as social network analysis, biological network modeling, and infrastructure network optimization. Overall, graph decomposition serves as a powerful tool to simplify, elucidate, and exploit the rich information embedded in complex networks. Basically, the proposed LO-HI decomposition method, a information-geometric network filtering algorithm, is composed of five major steps:

\begin{enumerate}
	\item Estimation of the inverse temperature parameter $\beta$, which is responsible for defining the global behavior (low $\beta$ means randomness and high $\beta$ means smoothness);
	\item Computation of the shape-operator based curvaure $S_i(\beta)$ at each node of the network;
	\item Normalize the values of $S_i(\beta)$ to the interval $[0, 1]$;
	\item Define the information threshold $T$ (i.e. the 75\% quantile);
	\item Define the two disjoint components of the decomposition:
		\begin{itemize}
			\item The \textbf{L-subgraph}, induced by the nodes having low information content, that is, nodes for which $S_i(\beta) < T$ (the patterns that are aligned to the expected behavior);
			\item The \textbf{H-subgraph}, induced by the nodes having high information content, that is, nodes for which $S_i(\beta) \geq T$ (the patterns that are not expected to happen for this value of $\beta$);
		\end{itemize}
\end{enumerate}

The main idea behind the proposed decomposition strategy is to perform a filtering procedure to the original network that mimics low-pass and high-pass filters in image processing: 1) in smooth random fields (large inverse temperature), the L-subgraph is an approximation of the original network and preserves most of the communinity structure, while the H-subgraph contains the details, that is, nodes belonging to edges and boundary regions; 2) in noisy random fields (small inverse temperature), the L-subgraph is mostly composed by nodes with coarse and rough neighboring patterns, while the H-subgraph is formed of nodes with smooth neighborhoods.

\section{Computational experiments and results}

In order to test and evaluate the proposed information-geometric network decomposition method, we performed two sets of computational experiments: the first one, with networks having low degree variability (k-NN graphs) built from multivariate datasets and the second one, with irregular networks (arbitrary degree distribution) from several network data repositories. To compare the communities from the original network and those obtained after the filtering and decomposition process, we analize four different metrics: conductance, modularity, coverage and performance \cite{Newman}. In the following, we briefly describe each metric:

\begin{itemize}
	\item Conductance is the ratio of the cost of a cut, which is the total weight of the edges that were cut, and smallest of the weights of a group induced by that cut. It is a measure of the relative ease with which information can flow between a partition and the rest of the network.
	\item Coverage of a partition is the ratio of the number of intra-community edges to the total number of edges in the network. High coverage values mean that more edges lie in communities than between communities, which is an indication of a good partition.
	\item Performance is the ratio of the number of intra-community edges plus inter-community non-edges with the total number of potential edges. High performance values reflect a community that is internally dense and externally sparse.
	\item Modularity measures the number of edges within communities and the number of edges between communities to a random graph with similar characteristics. High modularity values indicate that the number of edges within communities exceeds the number of edges between communities, which is an indication of a good partition.
\end{itemize}

In the first set of experiments, we selected 20 public classification datasets from the openML repository. The basic features of the networks used in the first set of experiments are shown in Table \ref{tab:data}. 

\begin{table}[htb]
	\centering
	\caption{Number of nodes, edges and groups (communities) of the k-NN graphs generated by each dataset ($k = 15$).}
	\begin{tabular}{ccccc}
		\toprule
		\textbf{\#} & \textbf{Datasets}  & \textbf{\# nodes} & \textbf{\# edges} & \textbf{\# communities} \\
		\midrule
		1 & iris   & 150	&   1431    & 3  \\
		2 & wine   & 178	&   1802    & 3  \\
		3 & breast\_cancer  & 569 & 6321  & 2  \\
		4 & digits  & 1797 & 18817  & 10  \\
		5 & prnn\_crabs &   200  & 1717  & 2  \\
		6 & smartphone   & 180  & 1924  & 6   \\
		7 & texture (20\%) & 1100 & 10764  & 11   \\
		8 & segment (20\%) & 1155 & 11183 & 7 \\
		9 & mfeat-karhunen & 2000 & 21192 & 10 \\
		10 & mfeat-pixel & 2000 & 20741 & 10 \\
		11 & mfeat-fourier & 2000  & 21610 & 10  \\
		12 & tecator & 240 & 2480 & 2  \\
		13 & satimage (20\%) & 1286 & 14124 & 6  \\
		14 & pendigits (20\%) & 1099 & 11011  & 10  \\		
		15 & optdigits (20\%) & 1124 & 11905 & 10 \\
		16 & MNIST\_784 (2.5\%) & 1750 & 20442 & 10 \\
		17 & Kuzushiji-MNIST (2.5\%) & 1750 & 22151 & 10 \\
		18 & USPS (2.5\%) & 1859 & 20428 & 10 \\
		19 & JapaneseVowels (20\%) & 1992 & 19962 & 9 \\
		20 & vowel & 990 & 9420 & 11 \\
		\bottomrule  
	\end{tabular}
	\label{tab:data}
\end{table} 

\begin{figure}
	\centering
	\includegraphics[scale=0.33]{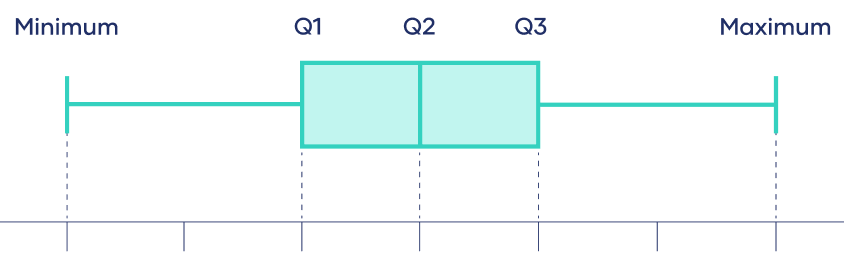}
	\caption{The high information nodes have value of $S_i(\beta)$ belonging to the interval $(Q3, max]$.}
	\label{fig:quartil}
\end{figure}

For each labeled network, we estimate the inverse temperature parameter and compute the value of $S_i(\beta)$ for each node. The threshold $T$ is computed as the value that defines the third quartile (Q3) of the distribution of the normalized $S_i(\beta)$. In other words, we define define $T$ as the 75\% percentile. Figure \ref{fig:quartil} shows a boxplot illustration that shows the minimum, Q1, Q2 (median), Q3 and the maximum.

After that, we compute the L-subgraph, as the subgraph induced by the L-nodes (below $T$) and the H-subgraph, as the subgraph induced by the H-nodes (above $T$). Finally, we compute the modularity, coverage and performance of the original network, L-subgraph and H-subgraph. Table \ref{tab:results1} summarizes the obtained results. By looking at the results, it is possible to see that the low information subgraph (L-subgraph) have the largest modularities and coverages, indicating that they improve the community structure of the original network. In other words, it can be thought as a smooth approximation, in the sense that it preserves most of the small range interactions (within-cluster edges) but removes most of the long range interactions (inter-cluster edges). On the other hand, the results indicate that the high information subgraph (H-subgraph) is composed mostly by long range interactions, building a representation that resembles the boundaries that interconnects the different communities. 

\begin{table}[htb]
	\centering
	\caption{Modularity, coverage and performance metrics for orignal network, L-subgraph and H-subgraph for several labeled k-NN graphs built from multivariate datasets ($k = 15$)}
	\begin{tabular}{cccccccccc}
		\toprule
		& \multicolumn{3}{c}{\textbf{Network}}     & \multicolumn{3}{c}{\textbf{L-subgraph}} & \multicolumn{3}{c}{\textbf{H-subgraph}}  \\
		\midrule
		\textbf{Dataset}          & \textbf{M} & \textbf{C} & \textbf{P}     & \textbf{M}     & \textbf{C}     & \textbf{P}     & \textbf{M}     & \textbf{C} & \textbf{P} \\
		\midrule
		\textbf{iris}             & 0.524      & 0.885      & 0.769          & \textbf{0.651} & \textbf{0.998} & \textbf{0.802} & 0.026          & 0.552      & 0.570      \\
		\textbf{wine}             & 0.513      & 0.891      & 0.751          & \textbf{0.641} & \textbf{0.999} & \textbf{0.790} & 0.002          & 0.767      & 0.402      \\
		\textbf{breast\_cancer}   & 0.408      & 0.921      & \textbf{0.501} & \textbf{0.495} & \textbf{1.000} & 0.499          & 0.176          & 0.683      & 0.523      \\
		\textbf{digits}           & 0.767      & 0.902      & \textbf{0.909} & \textbf{0.884} & \textbf{0.997} & \textbf{0.909} & 0.452          & 0.639      & 0.873      \\
		\textbf{prnn\_crabs}      & 0.157      & 0.707      & 0.538          & \textbf{0.413} & \textbf{0.928} & \textbf{0.599} & -0.019         & 0.534      & 0.516      \\
		\textbf{smartphone}       & 0.470      & 0.713      & 0.889          & \textbf{0.594} & \textbf{0.861} & \textbf{0.917} & 0.241          & 0.763      & 0.815      \\
		\textbf{texture}          & 0.688      & 0.838      & \textbf{0.920} & \textbf{0.852} & \textbf{0.984} & 0.915          & 0.311          & 0.526      & 0.857      \\
		\textbf{segment}          & 0.575      & 0.833      & \textbf{0.868} & \textbf{0.812} & \textbf{0.989} & 0.862          & 0.212          & 0.504      & 0.749      \\
		\textbf{mfeat-karhunen}   & 0.732      & 0.852      & \textbf{0.907} & \textbf{0.871} & \textbf{0.983} & \textbf{0.907} & 0.404          & 0.606      & 0.861      \\
		\textbf{mfeat-pixel}      & 0.809      & 0.924      & \textbf{0.909} & \textbf{0.893} & \textbf{0.999} & \textbf{0.909} & 0.455          & 0.629      & 0.876      \\
		\textbf{mfeat-fourier}    & 0.534      & 0.670      & \textbf{0.904} & \textbf{0.734} & \textbf{0.873} & 0.902          & 0.168          & 0.368      & 0.837      \\
		\textbf{tecator}          & 0.155      & 0.708      & \textbf{0.526} & \textbf{0.294} & \textbf{0.906} & 0.490          & -0.013         & 0.624      & 0.350      \\
		\textbf{satimage}         & 0.574      & 0.793      & \textbf{0.821} & \textbf{0.739} & \textbf{0.979} & 0.799          & 0.212          & 0.449      & 0.791      \\
		\textbf{pendigits}        & 0.698      & 0.862      & 0.913          & \textbf{0.872} & \textbf{0.991} & \textbf{0.917} & 0.403          & 0.611      & 0.874      \\
		\textbf{optdigits}        & 0.731      & 0.861      & 0.914          & \textbf{0.876} & \textbf{0.990} & \textbf{0.918} & 0.354          & 0.545      & 0.865      \\
		\textbf{MNIST\_784}       & 0.559      & 0.692      & 0.904          & \textbf{0.746} & \textbf{0.872} & \textbf{0.905} & 0.259          & 0.400      & 0.873      \\
		\textbf{Kuzushiji\_MNIST} & 0.458      & 0.576      & 0.901          & \textbf{0.640} & \textbf{0.756} & \textbf{0.905} & 0.170          & 0.283      & 0.881      \\
		\textbf{USPS}             & 0.674      & 0.825      & \textbf{0.900} & \textbf{0.837} & \textbf{0.977} & 0.899          & 0.376          & 0.522      & 0.880      \\
		\textbf{JapaneseVowels}   & 0.659      & 0.798      & \textbf{0.888} & \textbf{0.819} & \textbf{0.960} & 0.885          & 0.287          & 0.458      & 0.854      \\
		\textbf{vowels}           & 0.247      & 0.423      & 0.907          & \textbf{0.346} & \textbf{0.520} & \textbf{0.909} & 0.110          & 0.392      & 0.890      \\
		\midrule
		\textbf{Average}          & 0.547      & 0.784      & 0.827          & \textbf{0.700} & \textbf{0.928} & \textbf{0.832} & 0.229          & 0.543      & 0.757      \\
		\textbf{Median}           & 0.567      & 0.829      & 0.901          & \textbf{0.743} & \textbf{0.981} & \textbf{0.904} & 0.227          & 0.540      & 0.856      \\
		\textbf{Minimum}              & 0.155      & 0.423      & \textbf{0.501} & \textbf{0.294} & \textbf{0.520} & 0.490          & -0.019         & 0.283      & 0.350      \\
		\textbf{Maximum}              & 0.809      & 0.924      & \textbf{0.920} & \textbf{0.893} & \textbf{1.000} & 0.918          & 0.455          & 0.767      & 0.890      \\
		\bottomrule  
	\end{tabular}
	\label{tab:results1}
\end{table} 

\begin{figure}
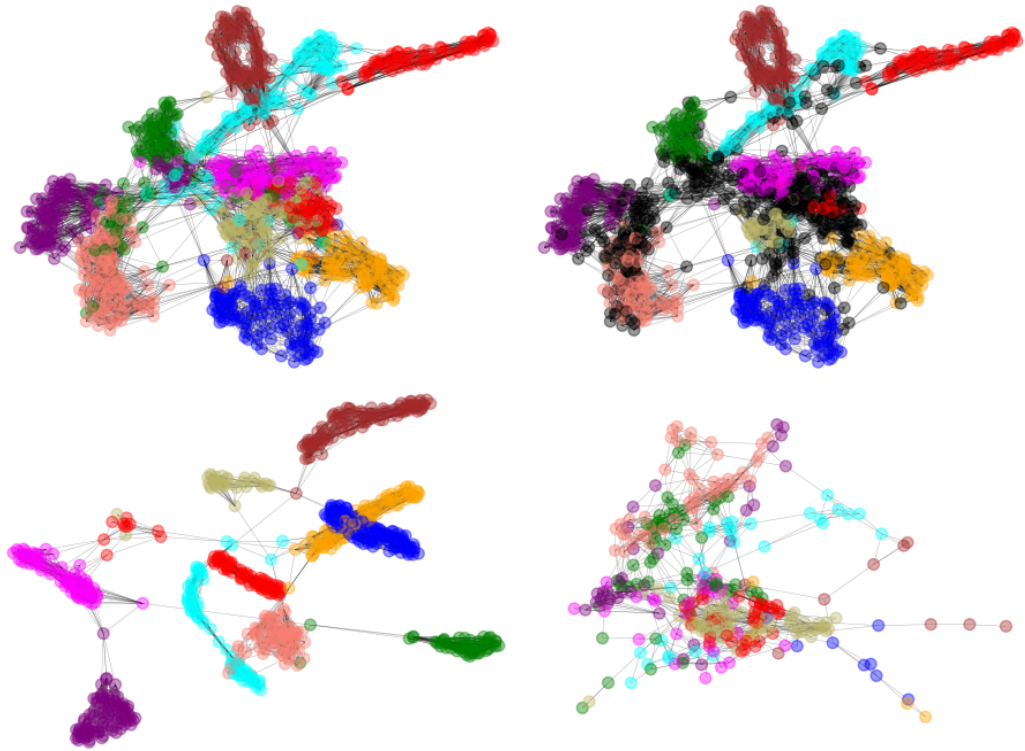

	\centering
	\includegraphics[scale=0.43]{pen\_comm.png}
	\includegraphics[scale=0.43]{pen\_comm\_black.png}
	\includegraphics[scale=0.43]{pen\_comm\_L.png}
	\includegraphics[scale=0.43]{pen\_comm\_H.png}
	\caption{Visualization of the qualitative results for the pendigits dataset. From left to right and top to bottom: a) Original network; b) Original network with high information nodes highlighted in black; c) L-subgraph (smooth); d) H-subgraph (boundaries).}
	\label{fig:mfeat}
\end{figure}

\begin{figure}
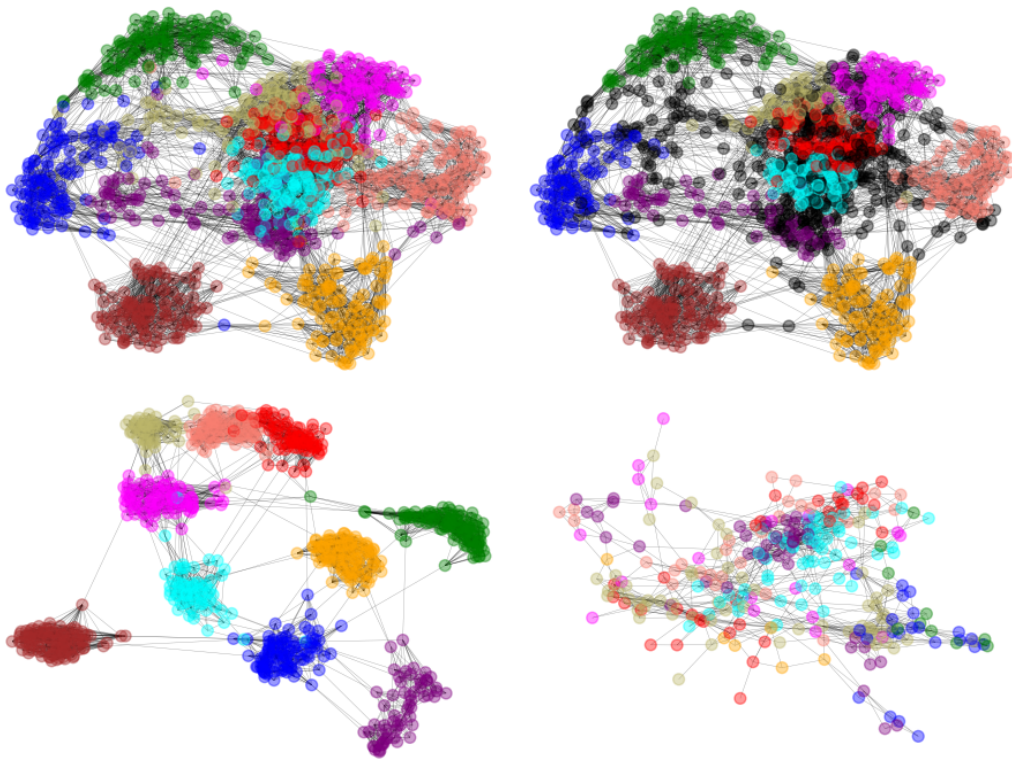

	\centering
	\includegraphics[scale=0.43]{opt\_comm.png}
	\includegraphics[scale=0.43]{opt\_comm\_black.png}
	\includegraphics[scale=0.43]{opt\_comm\_L.png}
	\includegraphics[scale=0.43]{opt\_comm\_H.png}
	\caption{Visualization of the qualitative results for the optdigits dataset. From left to right and top to bottom: a) Original network; b) Original network with high information nodes highlighted in black; c) L-subgraph (smooth); d) H-subgraph (boundaries).}
	\label{fig:opt}
\end{figure}

In order to show some visual representations of networks and their respective L-subgraph and H-subgraph, Figures \ref{fig:mfeat} and \ref{fig:opt} illustrate qualitative results for the pendigits and optdigits. Note that the L-subgraph is like a smooth version of the original network in the sense that modularity increases and the community structure becomes more evident without the boundary nodes. However, the visualization of the H-subgraph shows that the community structure is mostly absent, as the modularity is quite reduced. 

Another interesting finding is that the conductance between the L-nodes and H-nodes is large, indicating that the H-nodes have a relevant role in spreading information along the network communities due to their long range interactions (inter-communities). Table \ref{tab:results2} shows, for each network, a comparison between the largest conductance between a pair of communities in the original network and the conductance between the L-nodes and H-nodes. The partition induced by the proposed information-theoretic network filtering method seems to decompose the set of nodes in subsets that maximize the conductance. In other words, the method can be useful in applications that need to find nearly optimal conductance partitions, that is, the H-nodes are the ones responsible to spread information globally.

\begin{table}[htb]
	\centering
	\caption{Maximum conductance between pairwise communities in the original network and the conductance between L-nodes and H-nodes for several labeled k-NN graphs built from multivariate data ($k = 15$)}
	\begin{tabular}{ccc}
		\toprule
		\textbf{Dataset}          & \textbf{Pairwise communities} & \textbf{L-nodes/H-nodes} \\
		\midrule
		\textbf{iris}             & 0.167                  & \textbf{0.428}              \\
		\textbf{wine}             & 0.103                  & \textbf{0.536}              \\
		\textbf{breast\_cancer}   & 0.108                  & \textbf{0.352}              \\
		\textbf{digits}           & 0.069                  & \textbf{0.490}              \\
		\textbf{prnn\_crabs}      & 0.294                  & \textbf{0.305}              \\
		\textbf{smartphone}       & 0.277                  & \textbf{0.677}              \\
		\textbf{texture}          & 0.196                  & \textbf{0.424}              \\
		\textbf{segment}          & 0.184                  & \textbf{0.332}              \\
		\textbf{mfeat-karhunen}   & 0.167                  & \textbf{0.549}              \\
		\textbf{mfeat-pixel}      & 0.063                  & \textbf{0.515}              \\
		\textbf{mfeat-fourier}    & 0.392                  & \textbf{0.559}              \\
		\textbf{tecator}          & 0.362                  & \textbf{0.609}              \\
		\textbf{satimage}         & 0.253                  & \textbf{0.431}              \\
		\textbf{pendigits}        & 0.103                  & \textbf{0.420}              \\
		\textbf{optdigits}        & 0.091                  & \textbf{0.506}              \\
		\textbf{MNIST\_784}       & 0.182                  & \textbf{0.674}              \\
		\textbf{Kuzushiji\_MNIST} & 0.128                  & \textbf{0.631}              \\
		\textbf{USPS}             & 0.124                  & \textbf{0.555}              \\
		\textbf{JapaneseVowels}   & 0.129                  & \textbf{0.509}              \\
		\textbf{vowels}           & 0.308                  & \textbf{0.629}              \\
		\midrule
		\textbf{Average}          & 0.185                  & \textbf{0.507}     \\
		\textbf{Median}           & 0.167                  & \textbf{0.512}     \\
		\textbf{Minimum}              & 0.063                  & \textbf{0.305}     \\
		\textbf{Maximum}              & 0.392                  & \textbf{0.677}     \\
		\bottomrule  
	\end{tabular}
	\label{tab:results2}
\end{table}

In the second set of experiments, we use network data instead of k-NN graphs, where the key difference is the degree distribution. In k-NN graphs, the node degrees do not have significant variations, while in random networks, the variability of the node degrees is quite high. Often, in random networks, we observe few nodes with a very high degree and many nodes with very low degree. The impact of this variability in the proposed method is that, for some networks, the MPL estimator of the inverse temperature parameter can be severely overestimated \cite{MPL_over1,MPL_over2}. To overcome this limitation, we compute the critical inverse temperature. It has been shown that the value of the critical inverse temperature in the q-state Potts model is given by $\beta_c = \ln \left( 1 + \sqrt{q} \right)$ \cite{Critical,Critical2}. This equation is quite intuitive as the larger the number of states, the higher the inverse temperature required to cause a phase transition in the random field. Thus, in all the analysis along this second set of experiments, we define the estimative of the inverse temperature parameter as $\hat{\beta} = min\{ \beta_c, \beta_{MPL} \}$, where $\beta_{MPL}$ denotes the maximum pseudo-likelihood estimative. The basic features of the networks used in the second set of experiments are shown in Table \ref{tab:net}. 

\begin{table}[H]
	\centering
	\caption{Number of nodes, edges and groups (communities) of the networks.}
	\begin{tabular}{ccccc}
		\toprule
		\textbf{\#} & \textbf{Network}  & \textbf{\# nodes} & \textbf{\# edges} & \textbf{\# communities} \\
		\midrule
		1 & karate   & 34	&   78    & 3  \\
		2 & football   & 115	&  613   &  6 \\
		3 & dolphins  & 62 & 154 & 4 \\
		4 & Les miserables  & 77 & 254 & 5 \\
		5 & political books &  105 & 441 & 4 \\
		6 & soccer   & 35 & 118 & 5 \\
		7 & USAir97 & 332 & 2126 & 7 \\
		8 & bio-celegans & 453 & 2025 & 10 \\
		9 & bio-diseasome & 516 & 1188 & 24 \\
		10 & eco-everglades & 69 & 885 & 3 \\
		\bottomrule  
	\end{tabular}
	\label{tab:net}
\end{table}

\begin{figure}
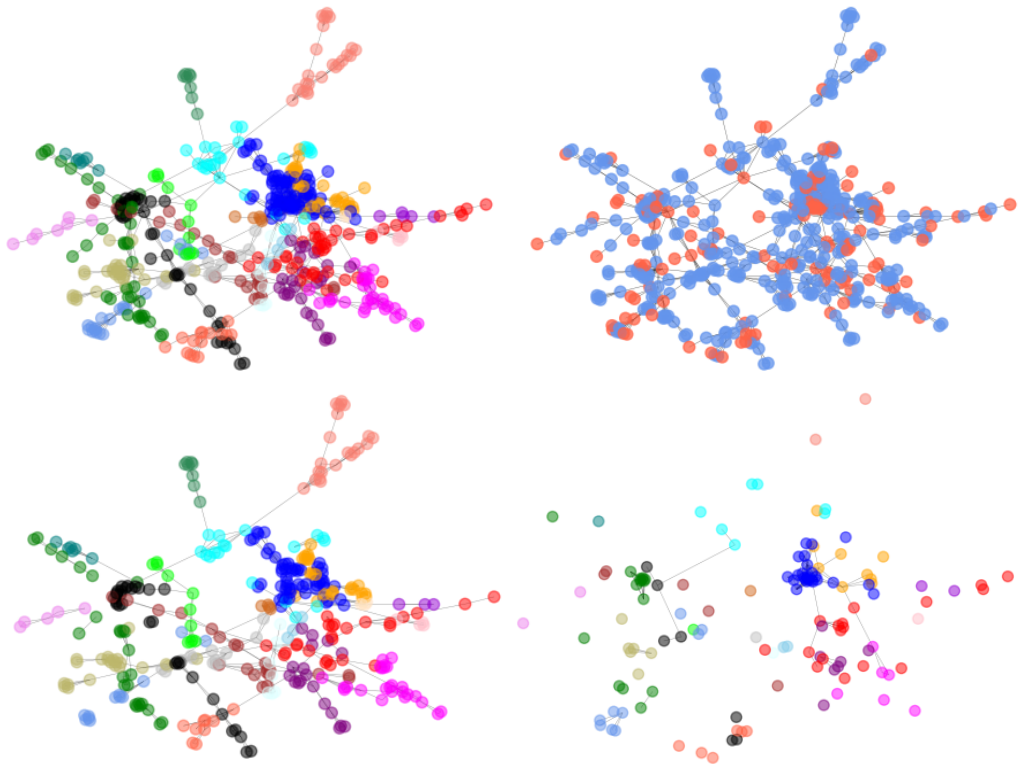

	\centering
	\includegraphics[scale=0.43]{dis\_comm.png}
	\includegraphics[scale=0.43]{dis\_high\_nodes.png}
	\includegraphics[scale=0.43]{dis\_L.png}
	\includegraphics[scale=0.40]{dis\_H.png}
	\caption{Visualization of the qualitative results for the bio-diseasome network. From left to right and top to bottom: a) Original network; b) Original network with high information nodes highlighted in red; c) L-subgraph (smooth); d) H-subgraph (boundaries).}
	\label{fig:dis}
\end{figure}

The methodology is similar to the one described in the first set of experiments: we apply the Clauset-Newman-Moore algorithm to detect the communities of the network and label its nodes, then the inverse temperature is computed and the information value $S_i(\beta)$ is evaluated for each node of the network. Finally, the LO-HI decomposition is performed and the modularity and coverage metrics are computed in both L-subgraph and H-subgraph. Table \ref{tab:results3} shows the obtained results. Once again, note that, in most cases, the L-subgraph increases both modularity and coverage, indicating that it is a smooth version of the original network that emphasizes the community structure. Figures \ref{fig:foot} and \ref{fig:dis} illustrate the results for the football and bio-diseasome networks.

\begin{table}[htb]
	\centering
	\caption{Modularity and coverage metrics for orignal network, L-subgraph and H-subgraph for several labeled networks}
	\begin{tabular}{ccccccc}
		\toprule
		& \multicolumn{2}{c}{\textbf{Original}} & \multicolumn{2}{c}{\textbf{L-subgraph}} & \multicolumn{2}{c}{\textbf{H-subgraph}} \\
		\midrule
		\textbf{Network}         & \textbf{M}          & \textbf{C}      & \textbf{M}         & \textbf{C}         & \textbf{M}       & \textbf{C}           \\
		\midrule
		\textbf{karate}          & 0.410               & 0.756           & \textbf{0.554}     & \textbf{0.961}     & 0.269            & 0.750                \\
		\textbf{football}        & 0.556               & 0.738           & \textbf{0.679}     & \textbf{0.874}     & 0.308            & 0.538                \\
		\textbf{dolphins}        & 0.495               & 0.823           & \textbf{0.503}     & \textbf{0.864}     & -0.125           & 0.500                \\
		\textbf{les miserables}  & 0.500               & 0.732           & \textbf{0.531}     & 0.787              & 0.000            & \textbf{1.000}       \\
		\textbf{political books} & \textbf{0.501}      & 0.918           & 0.497              & 0.888              & 0.460            & \textbf{0.967}       \\
		\textbf{soccer}          & 0.215               & 0.449           & \textbf{0.265}     & \textbf{0.606}     & -0.103           & 0.500                \\
		\textbf{USAir97}         & 0.144               & 0.769           & \textbf{0.445}     & \textbf{0.945}     & 0.062            & 0.758                \\
		\textbf{bio-celegans}    & 0.401               & 0.677           & \textbf{0.459}     & \textbf{0.698}     & 0.222            & 0.521                \\
		\textbf{bio-diseasome}   & 0.814               & 0.922           & \textbf{0.876}     & \textbf{0.940}     & 0.554            & 0.928                \\
		\textbf{eco-everglades}  & 0.111               & 0.541           & \textbf{0.150}     & \textbf{0.715}     & -0.033           & 0.362                \\
		\midrule
		\textbf{Average}         & 0.415               & 0.733           & \textbf{0.496}     & \textbf{0.828}     & 0.161            & 0.682                \\
		\textbf{Median}          & 0.453               & 0.747           & \textbf{0.500}     & \textbf{0.869}     & 0.142            & 0.644                \\
		\textbf{Minimum}             & 0.111               & 0.449           & \textbf{0.150}     & \textbf{0.606}     & -0.125           & 0.362                \\
		\textbf{Maximum}             & 0.814               & 0.922           & \textbf{0.876}     & 0.961              & 0.554            & \textbf{1.000}       \\
		\bottomrule  
	\end{tabular}
	\label{tab:results3}
\end{table} 

\begin{figure}
	\centering
	\includegraphics[scale=0.43]{football\_comm.png}
	\includegraphics[scale=0.43]{football\_high\_nodes.png}
	\includegraphics[scale=0.43]{football\_L.png}
	\includegraphics[scale=0.43]{football\_H.png}
	\caption{Visualization of the qualitative results for the football network. From left to right and top to bottom: a) Original network; b) Original network with high information nodes highlighted in red; c) L-subgraph (smooth); d) H-subgraph (boundaries).}
	\label{fig:foot}
\end{figure}

\begin{figure}
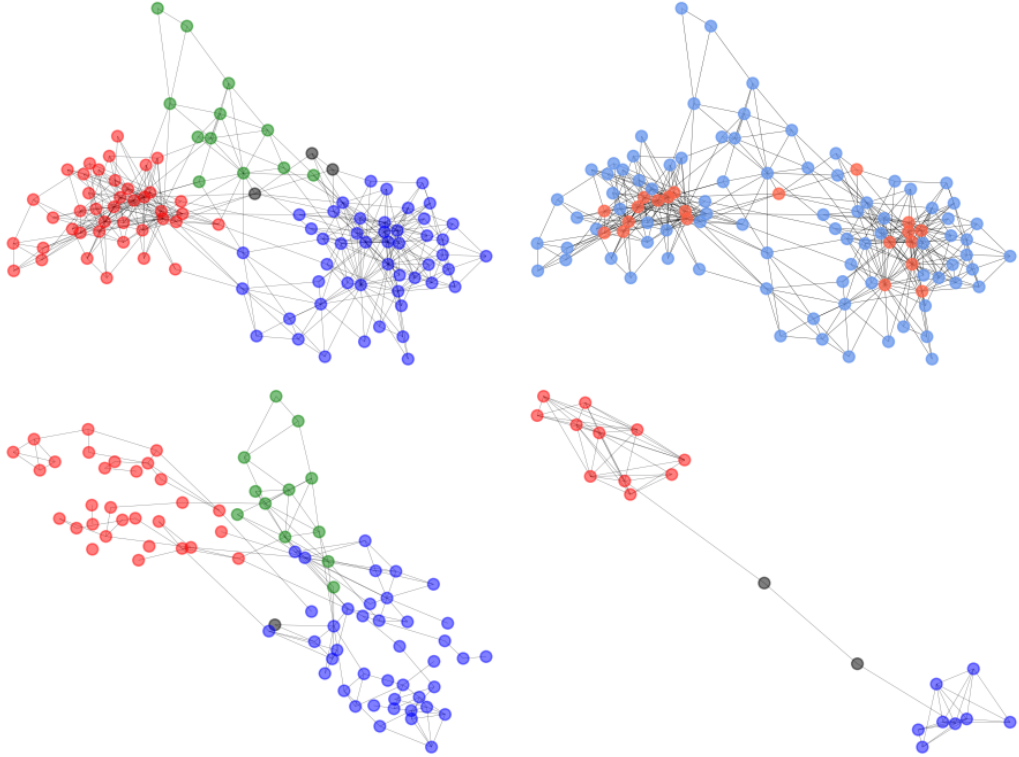

	\centering
	\includegraphics[scale=0.43]{books\_comm.png}
	\includegraphics[scale=0.43]{books\_high\_points.png}
	\includegraphics[scale=0.43]{books\_L.png}
	\includegraphics[scale=0.43]{books\_H.png}
	\caption{Visualization of the qualitative results for the political books network. From left to right and top to bottom: a) Original network; b) Original network with high information nodes highlighted in red; c) L-subgraph (low info); d) H-subgraph (high info).}
	\label{fig:books}
\end{figure}

A rather curious situation occurs in the analysis of the political books network, which is a network of books about US politics published around the time of the 2004 presidential election and sold by the online bookseller Amazon. Edges between books represent frequent copurchasing of books by the same buyers. As there is a large number of nodes with inter-community edges and few connections, the L-subgraph tends to keep those nodes, as this is the expected behavior. Hence, the H-subgraph is the component responsible to store the nodes that are not aligned to the expected behavior, which are those books deeply rooted in the core of the major communities (left and right of the political spectrum). In other words, what happens here is that the H-subgraph is smoother than the L-subgraph, contrary to all the other networks analyzed in this paper. Figure \ref{fig:books} shows an illustration of the L-subgraph and the H-subgraph. According to the results, the most informative books from the blue community are: Arrogance, Persecution, Off with their heads, A national party no more, Deliver us from evil, Dereliction of duty, Legacy and Losing Bin Laden. On the other side, the most informative books from the red community are: Against all enemies, The price of loyalty, Worse than Watergate, American dynasty, The great unraveling, Big lies, The lies of George W. Bush, Bushwhacked, Lies and the lying liars who tell them, Thieves in high places and Dude, where's my country? Hence, with the proposed LO-HI decomposition, it is possible to know the most and least informative nodes of a network, depending on its topology and overall community structure. FOr the interested reader, the Python source code to reproduce the results presented in this paper can be found at: \url{https://github.com/alexandrelevada/LOHI_decomposition}.

\section{Conclusions and final remarks}

Complex networks analysis holds significant relevance across a multitude of disciplines due to its ability to elucidate the intricate web of relationships and interactions inherent in various systems by uncovering the underlying structure and dynamics of complex systems, whether they are social, technological, biological, or informational in nature. Markov random fields (MRFs) provide a robust framework for modeling node interactions in complex networks by capturing the dependencies between nodes through a probabilistic approach. This framework assumes the Markov property, meaning that the state of a node is conditionally independent of all other nodes given its neighbors, allowing for the representation of complex dependencies and influences among nodes.

In this paper, we presented an innovative information-geometric approach for network filtering based on the q-state Potts model, a generalization of the Ising model to more than two states. Our method leverages the mathematical framework of information geometry to enhance the identification and extraction of significant structural features within complex networks. By applying the q-state Potts model, we model the network as a set of interacting nodes with multiple possible states, allowing for a nuanced representation of node interactions and community structures. This approach enables the filtering of noisy or irrelevant connections, thereby refining the network's topology and revealing its intrinsic organization. We demonstrate the efficacy of our method through both synthetic and real-world network data, showcasing its ability to uncover meaningful patterns and improve the interpretability of network structures. Our results highlight the potential of the q-state Potts model, coupled with information-geometric techniques, as a powerful tool for network analysis, offering significant advantages over traditional methods in terms of flexibility and robustness.

The research presented in this paper opens several avenues for future exploration and development. One immediate direction is the extension of the q-state Potts model to dynamically evolving networks. Incorporating temporal dynamics would allow our approach to handle real-time data streams and adapt to changes in network structure over time, thereby enhancing its applicability to domains such as social media analysis and biological network monitoring. Another promising area is the integration of multi-layer or multiplex networks, where nodes participate in multiple types of interactions across different layers. Extending our model to handle such complexity could provide deeper insights into systems where multiple types of relationships coexist, such as transportation networks, social networks with various forms of interactions, and biological systems with different types of molecular interactions. Further, investigating the theoretical underpinnings of the information-geometric approach in relation to other statistical mechanics models could yield a more comprehensive understanding of the model's capabilities and limitations. This could involve exploring different statistical ensembles and energy functions to enhance the model's flexibility and effectiveness in various contexts.

Future works may also include the development of more efficient computational algorithms for large-scale network filtering. The current implementation, while effective, may face scalability challenges with very large networks. Leveraging parallel computing and optimization techniques could significantly improve performance and enable the application of our method to massive datasets. Additionally, applying our approach to a broader range of real-world networks will be crucial for validating its generalizability and robustness. Case studies in diverse fields such as neuroscience, epidemiology, and infrastructure networks can help to identify domain-specific adaptations and optimizations that further refine the model. Lastly, user-friendly software tools and visualization frameworks that encapsulate our approach can facilitate its adoption by researchers and practitioners. Such tools would lower the barrier to entry and promote widespread use of the q-state Potts model for network filtering, fostering further advancements in the field. By pursuing these future directions, we aim to enhance the applicability, scalability, and impact of our information-geometric approach, contributing to the broader field of complex network analysis and fostering new discoveries across various domains.

\section*{Acknowledgments}
This study was partially funded by the Coordena\c{c}\~ao de Aperfei\c{c}oamento de Pessoal de N\'ivel Superior --- Brasil (CAPES), Finance Code 001.

\bibliographystyle{opticajnl}
\bibliography{sample}

\end{document}